\pdfoutput=1
\documentclass[lettersize,journal]{IEEEtran}
\usepackage{amsmath,amsfonts,amssymb}
\usepackage{array}
\usepackage[caption=false,font=footnotesize,labelfont=sf,textfont=sf]{subfig}
\usepackage{textcomp}
\usepackage{stfloats}
\usepackage{url}
\usepackage{verbatim}
\usepackage{graphicx}
\usepackage{cite}
\usepackage{booktabs}
\usepackage{tabularx}
\usepackage{pifont}
\usepackage[table]{xcolor}
\usepackage{multirow}
\usepackage[hidelinks]{hyperref}
\newcommand{\DoLP}{\mathrm{DoLP}}
\newcommand{\AoLP}{\mathrm{AoLP}}
\newcommand{\cmark}{\ding{51}}
\newcommand{\xmark}{\ding{55}}
\newcolumntype{L}{>{\raggedright\arraybackslash}X}
\newtheorem{lemma}{Lemma}
\newtheorem{proposition}{Proposition}

\begin{document}

\title{Polarization-Resolved Device Fingerprinting:\\ From Rotation-Encoded Stokes Imaging to a\\ Dual-Channel Metalens Reader}

\author{Alireza~Khalilian,~\IEEEmembership{Student Member,~IEEE,}
        Saad~Kiyani,
        Saeid~Khalilian,\\
        Hasan~Ali~Arif,
        and~Hafiz~Malik,~\IEEEmembership{Senior~Member,~IEEE}%
\thanks{This work has been submitted to the IEEE for possible publication. Copyright may be transferred without notice, after which this version may no longer be accessible.}%
\thanks{A. Khalilian, S. Kiyani, H. A. Arif, and H. Malik are with the Information Systems, Security and Forensics Lab, Department of Electrical and Computer Engineering, University of Michigan--Dearborn, Dearborn, MI 48128 USA (e-mail: akhalili@umich.edu; skiyani@umich.edu; hasanali@umich.edu; hafiz@umich.edu).}%
\thanks{S. Khalilian is with the Faculty of Computer Science and Mathematics, University of Passau, 94032 Passau, Germany.}%
\thanks{Preliminary versions of parts of this work were presented at the Conference on Lasers and Electro-Optics (CLEO), 2026, and at escar USA, 2026.}}%

\markboth{Preprint submitted to IEEE Transactions on Information Forensics and Security}%
{Khalilian \MakeLowercase{\textit{et al.}}: Polarization-Resolved Device Fingerprinting}

\maketitle

\begin{abstract}
Contactless authentication of electronic components requires a physical signature that can be read without powering the device. We study the linear polarization response of integrated-circuit markings as such a signature, using a rotation-encoded reference polarimeter at 450 and 660~nm and a reader without moving parts built from two polarization-selective silicon-rich-nitride metalenses at 660~nm, to our knowledge the first metalens-array reader for device fingerprinting. Least-squares Stokes estimates from 361-frame analyzer sweeps give a mean degree of linear polarization of 0.56--0.63 at 450~nm and 0.33--0.34 at 660~nm. With two independent half-sweeps as enrollment and probe, patch-level verification shows that polarization separates devices far more strongly than intensity. At 660~nm the decidability index rises from 1.5 for total intensity to 4.7 for the normalized Stokes vector and 13.0 for the angle of linear polarization, and fusing both wavelengths raises it to 19.5. We further show that enrollment does not transfer across wavelengths and that near-perfect closed-set accuracy is not by itself evidence of a surface fingerprint, because coarse shortcut features also reach it. For the metalens reader, the synthesized virtual-angle library is an exact re-parameterization of the two measured channels, and emulation on the reference data shows that at 660~nm a dual-channel snapshot doubles the decidability of the total intensity from the same frames and lowers the equal error rate from 3.1\% to 0.5\%. All results come from a single acquisition session per device, and we give the protocol that converts them into a cross-session test.
\end{abstract}


\section{Introduction}
\IEEEPARstart{E}{lectronics} supply chains are long and globally distributed, and counterfeit, recycled and remarked components remain a persistent risk for connected and safety-critical hardware~\cite{guin2014counterfeit,hammi2023security}. Labels, serial numbers and stored credentials can be copied or altered, which motivates authentication rooted in the physical characteristics of the part itself. Device fingerprinting distinguishes nominally identical units through variations introduced by materials, fabrication, packaging or hardware behavior~\cite{mir2025comprehensive,kumar2023device}, and supports counterfeit detection and supply-chain assurance in cyber-physical systems~\cite{wang2023fl4iot,ye2020cyber}.

Most device fingerprinting is \emph{signal based}, identifying a device from electrical, electromagnetic or communication signals observed while it operates~\cite{sanchez2024single,irfan2025device,chen2025physics,elmaghbub2024domain,radhakrishnan2014gtid}. These methods discriminate well under controlled conditions, but they require the part to be powered and active, and their performance depends on channel and operating conditions~\cite{chen2025physics,sanchez2024single}. Incoming inspection and forensic examination instead call for authentication of an unpowered part. \emph{Imaging-based} methods meet this need. Printed circuit boards (PCBs), printed structures, integrated-circuit (IC) packages and paper carry surface patterns that act as fingerprints~\cite{iqbal2017pcb,erozan2022counterfeit,dhanuskodi2020counterfoil,buchanan2005fingerprinting,sharma2011paperspeckle}. Intensity images, however, discard part of the information carried by light, and when genuine and counterfeit parts share geometry and markings, grayscale appearance becomes a weak discriminator.

\begin{figure*}[!t]
\centering
\includegraphics[width=\textwidth]{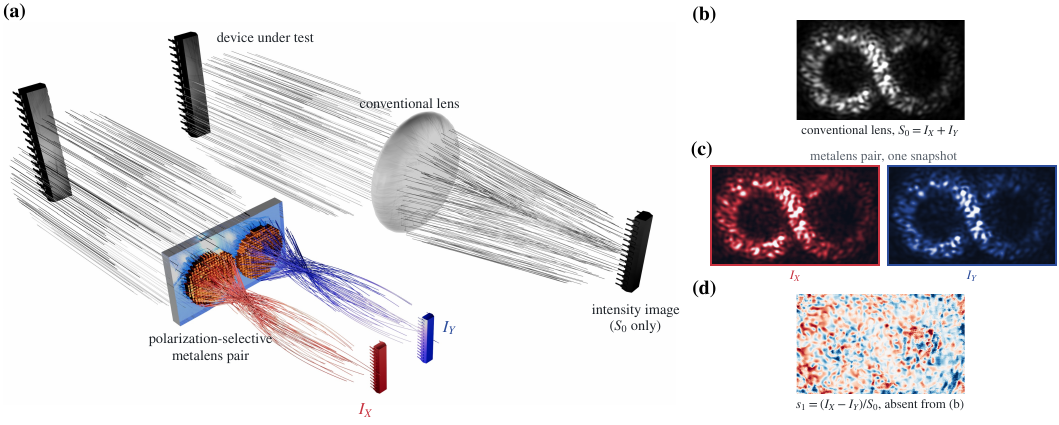}
\caption{How a polarization-selective metalens pair sees a device differently from a conventional lens. (a) Rendered comparison. A refractive lens images the device under test onto a single intensity image, while a pair of co-fabricated metalenses forms an $x$-polarized image $I_X$ (red) and a $y$-polarized image $I_Y$ (blue) in one snapshot. (b)--(d) Measured metalens data for one device at 660~nm. (b) Total intensity $S_0=I_X+I_Y$, which is what a conventional lens records. (c) The two polarization channels. (d) Normalized Stokes contrast $s_1=(I_X-I_Y)/S_0$, which cannot be obtained from (b).}
\label{fig:concept}
\end{figure*}

Polarization provides an additional, physically grounded channel. Reflection and scattering change the polarization state of light according to surface roughness, microstructure, stress, anisotropy and composition~\cite{wolff1990polarization,goldstein2011polarized}, so surfaces with similar intensity can have measurably different polarization responses~\cite{tyo2006review}. Polarization has been used against presentation attacks and counterfeit currency~\cite{abd2017face,santhanam2013counterfeit} and, as a property of radio signals, for LoRaWAN authentication~\cite{xu2023polarization}. Here it is an imaging observable of light--matter interaction at the device surface, available without powering the device. Practical polarimetry, however, usually relies on rotating analyzers or filter wheels, which add bulk, acquisition time and moving parts. Metasurfaces and metalenses integrate focusing and polarization selectivity in a single planar layer~\cite{yu2014flat,khorasaninejad2016metalenses} and have enabled compact Stokes cameras~\cite{arbabi2018full,rubin2019matrix} and single-shot computational imaging~\cite{li2024single}. Fig.~\ref{fig:concept} illustrates the difference. A pair of polarization-selective metalenses forms two orthogonally polarized images of a device marking in one exposure, from which a polarization contrast absent from a conventional intensity image can be computed.

Preliminary versions of this work introduced rotation-encoded polarization fingerprints of IC markings at two wavelengths~\cite{khalilian2026cleo} and a dual-channel metalens reader~\cite{khalilian2026escar}. To the best of our knowledge, the latter is the first use of a metalens array as the reader in device fingerprinting. Both reported near-perfect closed-set identification from a single acquisition session. A security evaluation must further establish whether the discriminative information is polarimetric or merely the layout of the marking, whether the accuracy survives protocols that prevent near-duplicate frames from leaking between training and test sets~\cite{kaufman2012leakage,roberts2017cross}, and how much polarization information a two-channel reader actually measures. This paper addresses these questions with a complete analysis of both datasets. The two readers play complementary roles. The rotation stage is a laboratory reference. Its 360 analyzer angles measure the full linear Stokes state, but each sweep takes 361 exposures, and the moving analyzer adds its own signature to the data. The metalens array needs a single exposure and has no moving parts, but it measures only two polarization states at one wavelength. The reference data test whether two states suffice, and at which wavelength. Our contributions are as follows.
\begin{enumerate}
\item A least-squares Stokes characterization of IC markings at 450 and 660~nm, with closed-form estimator noise that explains the limits of four-frame polarimetry (Sections~\ref{sec:theory}, \ref{sec:res-physics} and \ref{sec:res-fidelity}).
\item A verification protocol with inter-device registration and independent half-sweeps as enrollment and probe, which shows that polarization separates devices far more strongly than intensity, that enrollment is wavelength specific, and that the two wavelengths are complementary (Section~\ref{sec:res-verify}).
\item Leakage-aware identification protocols that respect the $180^\circ$ periodicity of Malus' law, with null and shortcut controls (Section~\ref{sec:res-ident}).
\item An analysis of the dual-channel metalens reader covering Stokes observability, the information content of the synthesized virtual-angle library and inter-channel registration (Sections~\ref{sec:reduced} and \ref{sec:res-metalens}).
\item A channel-count study that emulates two-, three- and four-channel snapshot polarimeters on the reference data and compares them with an intensity camera (Section~\ref{sec:res-channels}).
\end{enumerate}
We also state what a single acquisition session cannot establish and give the protocol that turns each experiment into a cross-session test (Section~\ref{sec:limits}).

\section{Related Work}
\subsection{Physical Fingerprints and Unclonable Functions}
Physical unclonable functions (PUFs) derive identity from uncontrollable fabrication variation~\cite{herder2014puf,yadav2022review}. Optical PUFs read the speckle of a disordered scattering token~\cite{pappu2002physical}, surface-texture fingerprints of paper apply the same principle to natural materials~\cite{buchanan2005fingerprinting,sharma2011paperspeckle}, and sensor pattern noise identifies cameras from their images~\cite{lukas2006digital}. For electronic components, counterfeit detection draws on electrical tests, physical inspection and imaging~\cite{guin2014counterfeit}, including PCB surface fingerprints~\cite{iqbal2017pcb}, optical signatures of printed structures~\cite{erozan2022counterfeit} and package-surface fingerprints of ICs captured with inexpensive cameras~\cite{dhanuskodi2020counterfoil}. Our work belongs to the imaging family but measures the linear polarization state rather than intensity.

\subsection{Polarimetric Imaging of Surfaces}
Reflected polarization depends on refractive index and surface microstructure, which underlies polarization-based material classification~\cite{wolff1990polarization}. Imaging polarimetry estimates \mbox{$S_0$--$S_2$} from analyzer projections, and the number and choice of analyzer states set the noise of the estimate~\cite{tyo2002design,tyo2006review,goldstein2011polarized}. Polarization cues have been used for face anti-spoofing~\cite{abd2017face} and currency inspection~\cite{santhanam2013counterfeit}, but not for the instance-level identity of nominally identical parts, and not with the evaluation pitfalls specific to angle-swept data in view.

\subsection{Metasurfaces in Imaging and Security}
Metasurfaces shape phase, amplitude and polarization with subwavelength meta-atoms~\cite{yu2014flat}. Metalenses focus at visible wavelengths with diffraction-limited performance~\cite{khorasaninejad2016metalenses}, and polarization-sorting metasurfaces enable full-Stokes imaging~\cite{arbabi2018full} and compact polarization cameras~\cite{rubin2019matrix}. Our reader uses polarization-selective silicon-rich-nitride metalenses~\cite{khalilian2025high,khalilian2026acs}. In security, metasurfaces have mainly served as the \emph{token}. Random plasmonic metasurfaces and nanostructured silicon labels act as optical PUFs read by dark-field microscopy or photoluminescence spectroscopy~\cite{kim2025quasi,wang2024allsilicon}. Metasurface polarization cameras have also reached commercial face-authentication modules for liveness detection~\cite{metalenz2023polarid}. To the best of our knowledge, a metalens array used as the polarimetric \emph{reader} that fingerprints unmodified electronic devices, as proposed in~\cite{khalilian2026escar} and analyzed here, has not been reported before.

\subsection{Evaluation Leakage}
When samples share acquisition context, random splits place near-duplicates in both training and test sets and inflate accuracy~\cite{kaufman2012leakage}. Blocked and grouped cross-validation are the standard remedies~\cite{roberts2017cross}. Angle-swept polarimetric stacks add the $180^\circ$ period of Malus' law, which a contiguous-block split does not respect.

\subsection{Positioning}
Table~\ref{tab:compare} compares representative approaches on five requirements for inspecting electronic components. Signal-based methods and electronic PUFs need an operating part, optical and metasurface PUFs need an attached token, surface fingerprints record intensity only, and polarization imaging has so far classified materials rather than identified units. Only the proposed approach meets all five requirements, and its metalens reader does so in one snapshot with a planar optic.

\begin{table*}[!t]
\caption{Comparison With Representative Fingerprinting and Authentication Approaches}\label{tab:compare}
\centering\footnotesize
\renewcommand{\arraystretch}{1.12}
\begin{tabularx}{\textwidth}{@{}>{\raggedright\arraybackslash}p{3.3cm}LLccccc@{}}
\toprule
Approach & Physical observable & Reader & \shortstack{Unpowered\\part} & \shortstack{Contact-\\less} & \shortstack{No design-in\\or tag} & \shortstack{Individual\\units} & \shortstack{Polarization-\\resolved}\\
\midrule
Signal-based device fingerprinting~\cite{kumar2023device,radhakrishnan2014gtid,elmaghbub2024domain} & Electrical, radio or timing behavior & Monitoring of device operation & \xmark & \cmark & \cmark & \cmark & \xmark\\
Radio polarization fingerprint~\cite{xu2023polarization} & Polarization of the radiated signal & Radio receiver & \xmark & \cmark & \cmark & \cmark & Radio only\\
Electronic PUFs~\cite{herder2014puf,yadav2022review} & On-chip process variation & Electrical challenge--response & \xmark & \xmark & \xmark & \cmark & \xmark\\
Optical PUF~\cite{pappu2002physical} & Speckle of a scattering token & Laser and camera & \cmark & \cmark & \xmark & \cmark & \xmark\\
Surface speckle~\cite{buchanan2005fingerprinting,sharma2011paperspeckle} & Surface microstructure (intensity) & Laser speckle imaging & \cmark & \cmark & \cmark & \cmark & \xmark\\
Surface imaging of parts~\cite{iqbal2017pcb,erozan2022counterfeit,dhanuskodi2020counterfoil} & Surface texture of PCBs, printed parts and IC packages (intensity) & Camera & \cmark & \cmark & \cmark & \cmark & \xmark\\
Metasurface PUF tags~\cite{kim2025quasi,wang2024allsilicon} & Random nanostructure of an added label & Dark-field microscopy or photoluminescence & \cmark & \cmark & \xmark & \cmark & \xmark\\
Polarization imaging~\cite{abd2017face,santhanam2013counterfeit,metalenz2023polarid} & Material-dependent reflection & Polarization imaging & \cmark & \cmark & \cmark & \xmark & \cmark\\
\midrule
\textbf{This work} & Linear Stokes field of the device marking & Rotating analyzer (reference) and planar dual-channel metalens array (snapshot) & \cmark & \cmark & \cmark & \cmark & \cmark\\
\bottomrule
\end{tabularx}
\end{table*}

\section{Threat Model and Problem Formulation}\label{sec:threat}
A \emph{verifier} (incoming-inspection station, service center or forensic laboratory) holds enrollment records for a set of genuine devices $\mathcal{K}=\{1,\dots,K\}$. Each record is a polarimetric fingerprint of a fixed region of interest (ROI), here the device marking. An unpowered part is imaged, and the verifier answers an \emph{identification} query (which enrolled device is this?) or a \emph{verification} query (is this the claimed device?). The \emph{adversary} substitutes a nominally identical part, such as a remarked, recycled or same-lot unit~\cite{guin2014counterfeit}, and can reproduce the layout, font and position of the marking, but not the sub-resolution surface structure that determines its local polarization response. The reader is trusted and the adversary does not control enrollment. Presentation attacks are discussed in Section~\ref{sec:discussion} but not evaluated, and physical cloning of the surface and attacks on the reader are outside the scope of this work.

\emph{Formulation.} Let $\mathbf F_k=\Phi(\mathcal M_k)$ denote the fingerprint that a reader $\Phi$ extracts from acquisition $\mathcal M_k$ of device $k$, and let $\mathbf T_k$ be the enrolled template. A verification query $(\mathbf F,k)$ is accepted when a similarity score exceeds a threshold,
\begin{equation}
\delta_\tau(\mathbf F,k)=\mathbf 1\!\left[s(\mathbf F,\mathbf T_k)\ge\tau\right].
\label{eq:decision}
\end{equation}
Genuine scores $s\sim f_{\mathrm g}$ arise from independent acquisitions of the same device, and impostor scores $s\sim f_{\mathrm i}$ from different devices that carry the same marking design. The false-match and false-non-match rates are $\mathrm{FMR}(\tau)=\Pr_{f_{\mathrm i}}[s\ge\tau]$ and $\mathrm{FNMR}(\tau)=\Pr_{f_{\mathrm g}}[s<\tau]$. The equal error rate (EER) is their common value where they cross, and the area under the receiver operating characteristic (ROC) curve (AUC) equals $\Pr[s_{\mathrm g}>s_{\mathrm i}]$ for independent draws. The security-relevant quantity is therefore not closed-set accuracy among a few enrolled units, but the separation of $f_{\mathrm g}$ and $f_{\mathrm i}$, which we summarize with the decidability index~\cite{daugman2004iris}
\begin{equation}
d'=\frac{\mu_{\mathrm g}-\mu_{\mathrm i}}{\sqrt{\tfrac12(\sigma_{\mathrm g}^2+\sigma_{\mathrm i}^2)}},
\label{eq:dprime}
\end{equation}
where $(\mu_{\mathrm g},\sigma_{\mathrm g})$ and $(\mu_{\mathrm i},\sigma_{\mathrm i})$ are the mean and standard deviation of the two score populations. Unlike the EER, which saturates at zero for small populations, $d'$ continues to rank representations after the empirical distributions separate.

\section{Theory}\label{sec:theory}
\subsection{Measurement Model}
Let $k\in\mathcal K$ index the device, $\lambda$ the wavelength and $\mathbf x=(x,y)\in\Omega_k$ a pixel in the ROI of device $k$. The light reflected by the marking at $\mathbf x$ is described by the Stokes vector $\mathbf S=(S_0,S_1,S_2,S_3)^{\mathsf T}$, which satisfies $S_0\ge\|(S_1,S_2,S_3)\|$. An ideal linear analyzer at angle $\theta$ has Mueller first row $\mathbf m(\theta)^{\mathsf T}=\tfrac12[1,\ \cos2\theta,\ \sin2\theta,\ 0]$, so the sensor records
\begin{equation}
I^{(k)}_\lambda(\mathbf x,\theta)=\mathbf m(\theta)^{\mathsf T}\mathbf S=\tfrac12\big[S_0+S_1\cos 2\theta+S_2\sin 2\theta\big]
\label{eq:malus}
\end{equation}
(indices on $\mathbf S$ omitted). Circular polarization $S_3$ is not observable with linear analyzers. Equation~(\ref{eq:malus}) is $\pi$-periodic in $\theta$, so frames at $\theta$ and $\theta+180^\circ$ are replicates of the same measurement, which matters for evaluation (Section~\ref{sec:protocol}).

\begin{figure*}[!t]
\centering
\includegraphics[width=\textwidth]{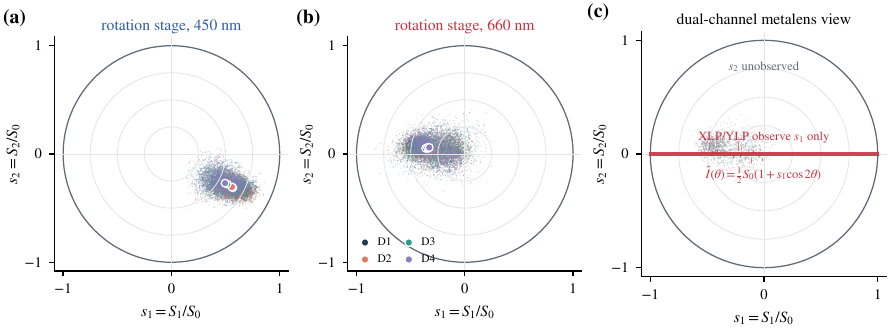}
\caption{Measured per-pixel linear Stokes vectors of the IC markings. (a), (b) Normalized Stokes vectors $(s_1,s_2)=(S_1,S_2)/S_0$ of 2500 random ROI pixels per device on the unit disk at 450 and 660~nm (large dots mark medians). The two wavelengths occupy nearly antipodal regions of the disk (polarization axes about $80^\circ$ apart). (c) A dual-channel reader with $x$- and $y$-linear-polarization (XLP/YLP) channels observes only the projection onto $s_1$. Every synthesized analyzer response $\hat I(\theta)$ is a function of $(S_0,s_1)$, and $s_2$ is not measured.}
\label{fig:disk}
\end{figure*}

For $N$ analyzer angles $\theta_0,\dots,\theta_{N-1}$, the samples of one pixel obey the linear model $\mathbf i=\mathbf A\mathbf p+\boldsymbol\eta$, with
\begin{equation}
\underbrace{\begin{bmatrix}I(\theta_0)\\ I(\theta_1)\\ \vdots\\ I(\theta_{N-1})\end{bmatrix}}_{\mathbf i}
=
\underbrace{\begin{bmatrix}1&\cos2\theta_0&\sin2\theta_0\\ 1&\cos2\theta_1&\sin2\theta_1\\ \vdots&\vdots&\vdots\\ 1&\cos2\theta_{N-1}&\sin2\theta_{N-1}\end{bmatrix}}_{\mathbf A}
\underbrace{\begin{bmatrix}a\\ b\\ c\end{bmatrix}}_{\mathbf p}
+\boldsymbol\eta,
\label{eq:linmodel}
\end{equation}
where $\mathbf p=\tfrac12(S_0,S_1,S_2)^{\mathsf T}$ and the noise satisfies $\mathbb E\boldsymbol\eta=\mathbf 0$ and $\operatorname{cov}\boldsymbol\eta=\sigma^2\mathbf I$. The least-squares estimate
\begin{equation}
\hat{\mathbf p}=(\mathbf A^{\mathsf T}\mathbf A)^{-1}\mathbf A^{\mathsf T}\mathbf i,\qquad \operatorname{cov}\hat{\mathbf p}=\sigma^2(\mathbf A^{\mathsf T}\mathbf A)^{-1},
\label{eq:ls}
\end{equation}
is the minimum-variance unbiased linear estimator (Gauss--Markov). From it we form the degree and angle of linear polarization and the normalized Stokes vector,
\begin{equation}
\begin{gathered}
\DoLP=\frac{\sqrt{S_1^2+S_2^2}}{S_0},\qquad
\psi=\AoLP=\tfrac12\operatorname{atan2}(S_2,S_1),\\
(s_1,s_2)=(S_1,S_2)/S_0,
\end{gathered}
\label{eq:dolp}
\end{equation}
so that $(s_1,s_2)=P\,(\cos2\psi,\sin2\psi)$ with $P=\DoLP$ lies in the unit disk (Fig.~\ref{fig:disk}). A small $\varepsilon$ is added to $S_0$.

\subsection{Estimator Precision and Analyzer Design}
The precision of (\ref{eq:ls}) depends on the analyzer set only through $\mathbf A^{\mathsf T}\mathbf A$. The following standard property covers every sampling scheme in this paper with $N\ge3$.

\begin{lemma}[Uniform analyzer frames]\label{lem:frame}
If $\theta_n=\theta_0+n\pi/N$, $n=0,\dots,N-1$, with $N\ge3$ (or the union of such sets), then $\mathbf A^{\mathsf T}\mathbf A=\operatorname{diag}(N,\,N/2,\,N/2)$. Hence $\hat a,\hat b,\hat c$ are uncorrelated and
\begin{equation}
\operatorname{var}(\hat a)=\frac{\sigma^2}{N},\qquad \operatorname{var}(\hat b)=\operatorname{var}(\hat c)=\frac{2\sigma^2}{N}.
\label{eq:var}
\end{equation}
\end{lemma}
\begin{IEEEproof}
The off-diagonal entries of $\mathbf A^{\mathsf T}\mathbf A$ and the deviations of its last two diagonal entries from $N/2$ are sums of $\cos(2m\theta_n)$ and $\sin(2m\theta_n)$ for $m\in\{1,2\}$, i.e., real and imaginary parts of $e^{\mathrm j2m\theta_0}\sum_n e^{\mathrm j2\pi mn/N}$. This geometric sum vanishes unless $N$ divides $m$, which is excluded for $N\ge3$.
\end{IEEEproof}

Under the conditions of Lemma~\ref{lem:frame}, (\ref{eq:ls}) reduces to discrete orthogonal projections onto the zeroth and second angular harmonics,
\begin{equation}
\begin{gathered}
\hat a=\frac1N\sum_{n=0}^{N-1}I_n,\\
\hat b=\frac2N\sum_{n=0}^{N-1}I_n\cos2\theta_n,\qquad
\hat c=\frac2N\sum_{n=0}^{N-1}I_n\sin2\theta_n,
\end{gathered}
\label{eq:proj}
\end{equation}
with $I_n=I(\mathbf x,\theta_n)$. Each Stokes map is thus a weighted sum of frames that can be accumulated during acquisition, and replacing $2\theta_n$ by $q\theta_n$ over the $360^\circ$ sweep gives the $q\theta$ harmonics of Section~\ref{sec:res-physics}. The quad ($0^\circ,45^\circ,90^\circ,135^\circ$) and triple ($0^\circ,60^\circ,120^\circ$) snapshot designs, a half-sweep ($N=180$, $1^\circ$ steps) and the full sweep ($N=360$) are all uniform frames in this sense and differ only through $N$. The dual design ($0^\circ,90^\circ$, $N=2$) is not a uniform frame. The third column of $\mathbf A$ vanishes, and $S_2$ is unobservable (Section~\ref{sec:reduced}). Lemma~\ref{lem:frame} is the equal-weighted-variance optimum of polarimeter design~\cite{tyo2002design} specialized to linear analyzers.

\begin{proposition}[Polarization-estimate noise]\label{prop:noise}
For a pixel with mean intensity $a$, degree of linear polarization $P$ and per-frame noise $\sigma$, under Lemma~\ref{lem:frame} and to leading order in $\sigma/(a\sqrt N)$,
\begin{equation}
\sigma_{\DoLP}\approx\frac{\sigma}{a}\sqrt{\frac{2+P^2}{N}},\qquad
\sigma_{\psi}\approx\frac{\sigma}{aP\sqrt{2N}},
\label{eq:dolpnoise}
\end{equation}
and the squared DoLP estimate is biased upward,
\begin{equation}
\mathbb E\big[\widehat{\DoLP}^{\,2}\big]\approx P^2+\frac{(4+3P^2)\,\sigma^2}{Na^2}.
\label{eq:bias}
\end{equation}
\end{proposition}
\begin{IEEEproof}
Write $r=\sqrt{b^2+c^2}$ and $\phi=\operatorname{atan2}(c,b)=2\psi$, so that $\DoLP=r/a$. By Lemma~\ref{lem:frame} the covariance of $(\hat b,\hat c)$ is isotropic, $(2\sigma^2/N)\mathbf I$, and the gradients of $r$ and $\phi$ with respect to $(b,c)$ are orthogonal with norms $1$ and $1/r$. First-order propagation therefore gives $\operatorname{var}(\hat r)\approx2\sigma^2/N$ and $\operatorname{var}(\hat\phi)\approx2\sigma^2/(Nr^2)$. Combining $\operatorname{var}(\hat r)/a^2$ with $r^2\operatorname{var}(\hat a)/a^4$ yields $\sigma_{\DoLP}$, and $\operatorname{var}(\hat\psi)=\operatorname{var}(\hat\phi)/4$ yields $\sigma_\psi$. For the bias, $\mathbb E[\hat b^2+\hat c^2]=r^2+4\sigma^2/N$, and multiplying by $\mathbb E[\hat a^{-2}]\approx a^{-2}\big(1+3\sigma^2/(Na^2)\big)$, with $\hat a$ independent of $(\hat b,\hat c)$, gives (\ref{eq:bias}). Section~S-I of the supplementary material gives the full derivation.
\end{IEEEproof}

Three consequences guide the experiments. First, the four-frame estimator used in~\cite{khalilian2026escar} ($N=4$) is noisier than a 360-frame fit by a factor of $\sqrt{90}\approx9.5$ in standard deviation, and $B\times B$ binning divides $\sigma$ by $B$ for uncorrelated pixel noise (Section~\ref{sec:res-fidelity}). Second, the orientation $\psi$ is least reliable where the pixel is weakly polarized, so AoLP should be read jointly with $P$. Third, by (\ref{eq:bias}) weakly polarized regions acquire spurious DoLP at small $N$, which makes few-frame DoLP maps look more alike across devices than they are. The linearization requires $r\gg\sigma\sqrt{2/N}$. Below this level the Rician bias of $\hat r$ dominates.

\subsection{Reduced Dual-Channel Polarimetry}\label{sec:reduced}
A polarization-selective metalens pair images the target through orthogonal analyzers~\cite{khalilian2025high} and records
\begin{equation}
\begin{bmatrix}I_X\\ I_Y\end{bmatrix}=\mathbf W\begin{bmatrix}S_0\\ S_1\\ S_2\end{bmatrix},\qquad
\mathbf W=\frac12\begin{bmatrix}1&\phantom{-}1&0\\ 1&-1&0\end{bmatrix},
\label{eq:dual}
\end{equation}
so that $S_0=I_X+I_Y$ and $S_1=I_X-I_Y$. The null space of $\mathbf W$ is the $S_2$ axis, so two polarization states that differ only in $S_2$ produce identical captures. The reduced contrast is
\begin{equation}
\widetilde{\DoLP}=\frac{|S_1|}{S_0}=P\,|\cos2\psi|\;\le\;\DoLP,
\label{eq:reduced}
\end{equation}
with equality when the polarization is aligned with the reader axes ($\psi\in\{0^\circ,90^\circ\}$) and zero at $\psi=\pm45^\circ$. A dual-channel reader is thus a projection of the Stokes disk onto the $s_1$ axis [Fig.~\ref{fig:disk}(c)], and its sensitivity depends on the orientation of the reader relative to the mean AoLP of the device population.

The virtual-angle response used in~\cite{khalilian2026escar},
\begin{equation}
\hat I(\theta)=I_X\cos^2\theta+I_Y\sin^2\theta=\tfrac12\big[S_0+S_1\cos2\theta\big],
\label{eq:virtual}
\end{equation}
is (\ref{eq:malus}) with $S_2$ set to zero. Evaluated over a set of virtual angles $\Theta$, it yields the library $\mathcal L^{(k)}_\Theta=\{\hat I^{(k)}(\theta)\}_{\theta\in\Theta}$, whose information content is fixed by (\ref{eq:dual}).

\begin{proposition}[Rank of the virtual library]\label{prop:rank}
For any angle set $\Theta$, the images of $\mathcal L^{(k)}_\Theta$ lie in $\operatorname{span}\{S_0,S_1\}$, and after subtracting their mean over $\Theta$ the stack has rank at most one. Consequently, every feature computed from the library is a function of $(I_X,I_Y)$, and a train/test split across virtual angles separates deterministic transformations of the same two images rather than independent observations.
\end{proposition}
\begin{IEEEproof}
By (\ref{eq:virtual}), $\hat I(\theta)=\tfrac12S_0+\tfrac12\cos2\theta\,S_1$. Subtracting the mean over $\Theta$ leaves $\tfrac12(\cos2\theta-\overline{\cos2\theta})\,S_1$, a scalar multiple of the single image $S_1$.
\end{IEEEproof}

\subsection{Patch-Level Verification Score}
The marking is a deterministic pattern, so a meaningful impostor comparison must align the impostor's marking to the enrolled one. Otherwise, misregistration alone lowers impostor scores and inflates separation. We register every device to a common reference with an enhanced-correlation-coefficient (ECC) affine warp $\mathcal W_k$~\cite{evangelidis2008ecc}, estimated on $S_0$ and applied to all representations of that device. A representation is a $C$-channel map $\mathbf F\in\mathbb R^{C\times n_y\times n_x}$ ($C=1$ for $S_0$ and DoLP, and $C=2$ for $(s_1,s_2)$ and for AoLP encoded as $(\cos2\psi,\sin2\psi)$). For a patch $p$ of a $G_y\times G_x$ grid, let $\tilde{\mathbf F}_p$ be the patch with each channel centered to zero mean. The score is the largest correlation over local shifts $\boldsymbol\delta$,
\begin{equation}
s_p(\mathbf F,\mathbf F')=\max_{\|\boldsymbol\delta\|_\infty\le\Delta}\;
\frac{\langle\tilde{\mathbf F}_p,\,\tilde{\mathbf F}'_{p+\boldsymbol\delta}\rangle}{\|\tilde{\mathbf F}_p\|\,\|\tilde{\mathbf F}'_{p+\boldsymbol\delta}\|}.
\label{eq:score}
\end{equation}
Channel-wise centering removes the global polarization state shared by all devices, so the score responds to the \emph{spatial} structure of the polarization field. The maximization raises impostor scores more than genuine ones but also narrows their spread, so it can change $d'$ in either direction. The same $\Delta$ and alignment are applied to both populations.

\emph{Multi-wavelength fingerprints.} With a wavelength set $\Lambda=\{\lambda_1,\lambda_2,\dots\}$, the fingerprint of device $k$ is the collection $\mathbf F^{(k)}=\{\mathbf F^{(k)}_{\lambda}\}_{\lambda\in\Lambda}$ of per-wavelength representations. Because the polarization axis of the marking changes with wavelength (Section~\ref{sec:res-physics}), the components are compared wavelength by wavelength and the patch scores are fused with the sum rule,
\begin{equation}
s^{\Lambda}_p(\mathbf F,\mathbf F')=\frac{1}{|\Lambda|}\sum_{\lambda\in\Lambda}s_p\big(\mathbf F_\lambda,\mathbf F'_\lambda\big).
\label{eq:fusion}
\end{equation}

\section{Experimental Systems and Datasets}\label{sec:systems}
\subsection{Rotation-Encoded Reference Polarimeter}
\begin{figure*}[!t]
\centering
\includegraphics[width=\textwidth]{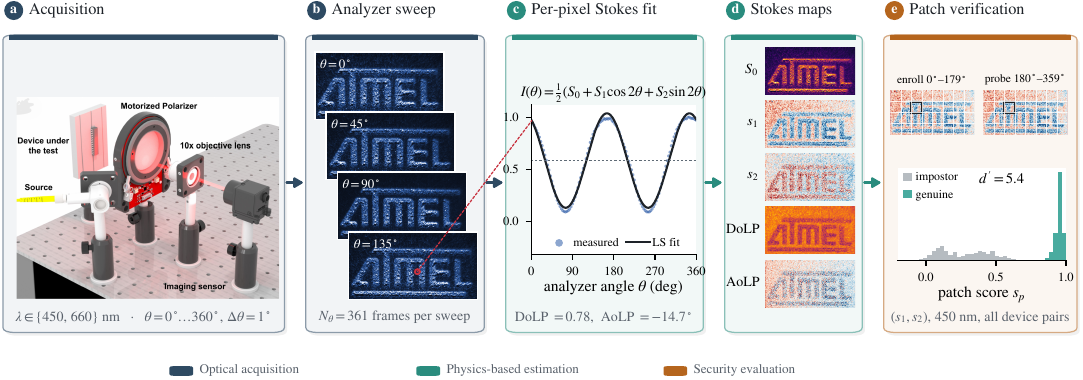}
\caption{Rotation-encoded reference polarimeter and processing pipeline with measured data (device D1, 450~nm). Card colors mark optical acquisition (dark blue), physics-based estimation (teal) and security evaluation (amber). (a) Reflected light from the device under test passes a motorized linear polarizer and a $10\times$ objective onto the imaging sensor. (b) Registered frames at four of the 361 analyzer angles. (c) Intensity of the marked pixel ($3\times3$ mean) against analyzer angle (dots, every third sample), the least-squares (LS) fit of (\ref{eq:malus}) and $S_0/2$ (dashed). (d) Stokes maps from the fit, with signed maps displayed about their median. (e) Independent half-sweeps as enrollment and probe, shown as $s_1$ maps with the $8\times4$ patch grid (one patch outlined), and the genuine and impostor score distributions of $(s_1,s_2)$ over all device pairs at 450~nm.}
\label{fig:setup}
\end{figure*}

The reference system (Fig.~\ref{fig:setup}) illuminates the device under test in reflection with fiber-coupled monochromatic sources at 450 and 660~nm.%
{} The reflected light passes a motorized linear polarizer and a $10\times$ objective onto a CMOS sensor. For each device and wavelength the analyzer is rotated from $0^\circ$ to $360^\circ$ in $1^\circ$ steps, giving $N_\theta=361$ frames. The $360^\circ$ frame duplicates $0^\circ$ and is excluded from all fits, so the full sweep has $N=360$. Frames are registered to a reference frame by ECC alignment and cropped to a $538\times1024$-pixel ROI containing the ATMEL marking. Four nominally identical ICs are used (D1--D4), and frames are box-averaged to $134\times256$ pixels unless stated otherwise. Image sizes are given as width $\times$ height, and maps are displayed rotated to landscape. A precomputed DoLP sequence (206 images per device and wavelength) is used only in the identification experiments.

\subsection{Dual-Channel Metalens Reader}
\begin{figure*}[!t]
\centering
\includegraphics[width=\textwidth]{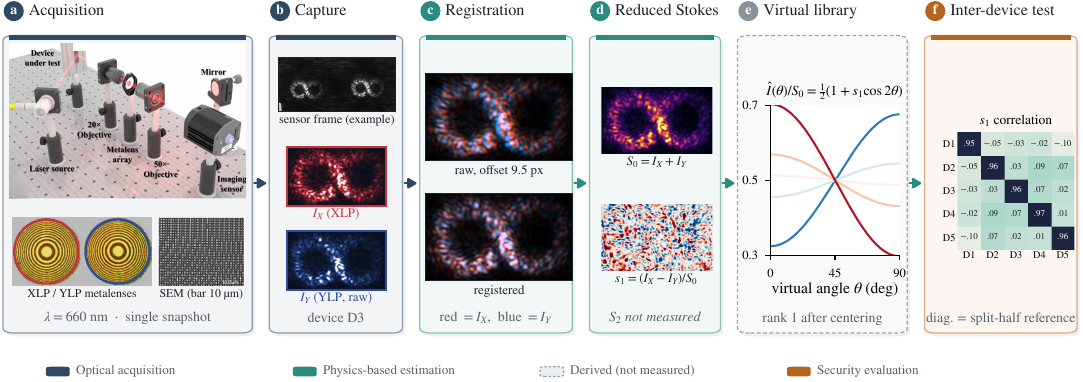}
\caption{Dual-channel metalens reader and processing pipeline at 660~nm with measured data (device D3 unless noted). Card colors follow Fig.~\ref{fig:setup}, and the dashed card marks derived data. (a) Imaging configuration, optical micrograph of the two co-fabricated polarization-selective metalenses (XLP red, YLP blue) and scanning electron microscope (SEM) image of the meta-atom array (scale bar 10~\textmu m). (b) Sensor frame with both lens images of the infinity-logo target, and the extracted $I_X$ and raw $I_Y$ channels. (c) Overlay of $I_X$ (red) and $I_Y$ (blue) before and after software registration of the YLP channel (raw offset 9.5~pixels). (d) Reduced Stokes maps $S_0$ and $s_1$ ($S_2$ is not measured). (e) Virtual-angle library (\ref{eq:virtual}), derived from (d) and of rank one after mean subtraction (Proposition~\ref{prop:rank}). Curves show $\hat I(\theta)/S_0$ for pixels at the 5th--95th percentiles of $s_1$. (f) Correlation of registered $s_1$ maps between devices. The diagonal correlates interleaved pixel rows of one image (a split-half reference set by pixel noise, not repeatability).}
\label{fig:metalens-hw}
\end{figure*}

The compact reader (Fig.~\ref{fig:metalens-hw}) replaces the rotating analyzer with a pair of co-fabricated, polarization-selective silicon-rich-nitride metalenses designed for 660~nm. Each lens imposes the hyperboloidal focusing phase
\begin{equation}
\phi_L(x,y)=\frac{2\pi}{\lambda}\Big(f-\sqrt{f^2+x^2+y^2}\Big),
\label{eq:metalens_phase}
\end{equation}
with focal length $f$ at the design wavelength $\lambda$, realized by distinct birefringent meta-atom distributions for the lenses selective to $x$- and $y$-linear polarization (XLP and YLP). The design and fabrication are reported in~\cite{khalilian2025high,khalilian2026acs}, and only the functional imaging model (\ref{eq:dual}) is used here. The 660~nm operating point suits a compact reader. Red AlGaInP laser diodes at this wavelength are a mature technology developed for recordable DVD drives~\cite{yagi2003high}, so low-cost sources are widely available, and on the reference data a dual-channel snapshot improves on intensity only at 660~nm (Section~\ref{sec:res-channels}). A laser source illuminates the target, a $20\times$ objective relays it onto the metalens array, and a $50\times$ objective images the two focal fields onto the sensor, producing XLP and YLP images of the same target region in one exposure. Five devices carrying an infinity-logo marking are imaged, each with one XLP and one YLP capture ($389\times230$~pixels). The processed dataset contains a contrast-normalized XLP image, a YLP image registered to it in software, and a 400-frame virtual-angle library at non-uniform angles between $0.15^\circ$ and $89.92^\circ$. Table~\ref{tab:data} summarizes both datasets.

\begin{table}[!t]
\caption{Datasets}\label{tab:data}
\centering\footnotesize
\begin{tabular}{@{}lll@{}}
\toprule
 & Rotation stage & Metalens reader\\
\midrule
Target & ATMEL IC marking & Infinity logo\\
Devices & 4 (D1--D4) & 5 (D1--D5)\\
Wavelength & 450, 660~nm & 660~nm\\
Analyzer settings & 361 angles ($1^\circ$ steps) & 2 (XLP, YLP)\\
Measured images & $4\times2\times361=2888$ & $5\times2=10$\\
Derived images & 206 DoLP / device / $\lambda$ & 400 virtual / device\\
ROI (pixels) & $538\times1024$ & $389\times230$\\
Sessions & 1 per device & 1 per device\\
\bottomrule
\end{tabular}
\end{table}

\section{Evaluation Protocol}\label{sec:protocol}
\subsection{Physics-Level Verification}
Each sweep contains two full periods of Malus' law. We fit (\ref{eq:ls}) separately to the first half-sweep ($0^\circ$--$179^\circ$) and to the second ($180^\circ$--$359^\circ$). The two halves are acquired at different times through the same optical path and are therefore independent measurements of the same polarization field within one mount. One serves as enrollment and the other as probe, in both directions. We compare five representations, namely a single analyzer frame $I(\theta_0)$ with $\theta_0=0^\circ$ (an intensity camera behind a fixed polarizer), $S_0$, DoLP, the normalized Stokes vector $(s_1,s_2)$ and AoLP encoded as $(\cos2\psi,\sin2\psi)$. Scores use (\ref{eq:score}) with an $8\times4$ ($G_y\times G_x$) patch grid and $\Delta=4$~pixels at $134\times256$ resolution, giving 256 genuine and 768 impostor scores per representation and wavelength. For the cross-wavelength test, enrollment uses the 450~nm half-sweep and the probe the 660~nm half-sweep, registered per device.

\subsection{Identification Protocols}
For comparison with~\cite{khalilian2026cleo}, we also evaluate closed-set identification on frame-level features. Because $I(\theta)=I(\theta+180^\circ)$ under (\ref{eq:malus}), a split that holds out the last 30\% of a $0^\circ$--$360^\circ$ sweep still leaves the $180^\circ$ twin of each test frame in the training set. We measure leakage with the $\pi$-periodic angular distance
\begin{equation}
d_\pi(\theta,\theta')=\min_{m\in\mathbb Z}\,|\theta-\theta'+m\,180^\circ|,
\label{eq:dpi}
\end{equation}
and call a split \emph{angle-disjoint} with guard $g$ if $d_\pi(\theta,\theta')>g$ for every training angle $\theta$ and test angle $\theta'$. We compare four protocols, namely (i) a random frame split, (ii) the contiguous-block split used previously (last 30\% of angles held out), (iii) the same test set after removing every training frame within $5^\circ$ of a test angle modulo $180^\circ$ (angle-disjoint with $g=5^\circ$), and (iv) three-fold cross-validation over $60^\circ$ sectors modulo $180^\circ$ with a $5^\circ$ guard band. For the DoLP sequence the corresponding period is $90^\circ$. Each frame $\mathcal M$ is box-averaged to $34\times64$ pixels and mapped to the descriptor
\begin{equation}
\mathbf z=h(\mathcal M)=\mathbf U^{\mathsf T}\boldsymbol\Sigma^{-1}\big(\operatorname{vec}\mathcal M-\boldsymbol\mu\big)\in\mathbb R^{d},
\label{eq:encoder}
\end{equation}
where $\boldsymbol\mu$ and $\boldsymbol\Sigma=\operatorname{diag}(\boldsymbol\sigma)$ standardize each pixel and $\mathbf U$ holds the first $d=64$ principal directions, all estimated on the training frames only. The device label is predicted by one-versus-rest ridge regression,
\begin{equation}
\hat y=\arg\max_{k\in\mathcal K}\;\mathbf w_k^{\mathsf T}\mathbf z+\beta_k,
\label{eq:classifier}
\end{equation}
with $(\mathbf w_k,\beta_k)$ fitted to $\pm1$ targets with penalty $\gamma=1$. We deliberately use a transparent linear model so that the protocol, not the feature extractor, determines the outcome. An optional ResNet-18~\cite{he2016deep} pipeline is provided with the code.

\subsection{Controls}
\emph{Null controls} run within one device. Random labels on random frames estimate chance, and labels that mark four contiguous angle blocks test whether a random frame split lets the classifier separate positions within a single sweep, which would indicate near-duplicate leakage. \emph{Shortcut controls} restrict the features to the image mean and standard deviation or to an 8-pixel ($2\times4$) block-average thumbnail of the ROI, and repeat identification with brightness-normalized frames, to test whether coarse acquisition factors alone identify the device.

\section{Results}\label{sec:results}
\subsection{Polarized, Wavelength-Dependent Marking Response}\label{sec:res-physics}
\begin{figure*}[!t]
\centering
\includegraphics[width=\textwidth]{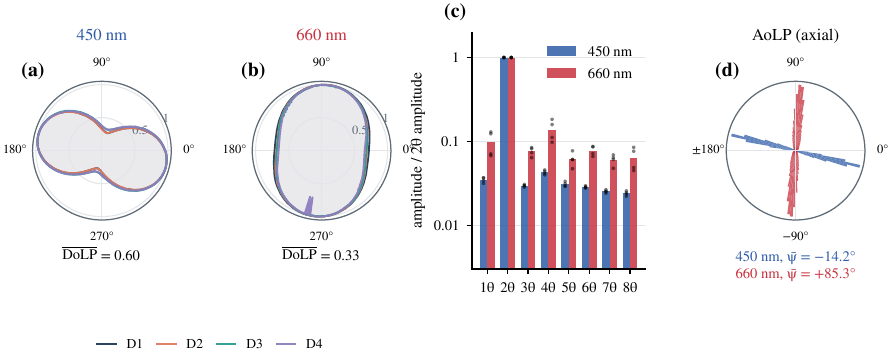}
\caption{Physics of the rotation-encoded signal. (a), (b) Mean intensity of the brightest 30\% of ROI pixels vs.\ analyzer angle on polar axes, normalized per device. The figure-eight is the $\cos^2$ law of linear polarization. (c) Harmonic amplitudes of the $360^\circ$ sweep relative to the $2\theta$ line (device means as bars, devices as dots). (d) DoLP-weighted AoLP histogram drawn as an axial rose (radius $\propto$ square root of the weighted count). The two inward spikes of D4 at 660~nm in (b) are single dark frames ($256^\circ$ and $259^\circ$, about 28\% below the model). They are retained in all analyses and change the half-sweep $S_0$ of these pixels by 0.4\%.}
\label{fig:signal}
\end{figure*}

Fig.~\ref{fig:signal}(a) and (b) show the mean intensity of the brightest ROI pixels against analyzer angle on polar axes. Both wavelengths produce the two-lobed figure of Malus' law, with a much narrower waist at 450~nm, where the mean per-pixel DoLP is 0.56--0.63 across devices, than at 660~nm, where it is 0.33--0.34. The harmonic content of the sweep [Fig.~\ref{fig:signal}(c)] confirms the model (\ref{eq:malus}). At 450~nm the $2\theta$ term explains 96.5--97.7\% of the per-pixel variance and the $1\theta$ and $4\theta$ lines stay below 5\% of the $2\theta$ line, whereas at 660~nm the $2\theta$ term explains 79.7--92.0\% and the $1\theta$ and $4\theta$ lines reach 13\% and 18\%. A $1\theta$ component cannot arise from an ideal analyzer. It indicates an image displacement that follows the rotation of the polarizer (beam wobble), while the $4\theta$ component indicates polarization-dependent transmission behind the analyzer or residual misregistration. Both arise from rotating the analyzer, which a snapshot reader avoids.

The polarization axis is tightly concentrated and wavelength dependent [Fig.~\ref{fig:signal}(d)]. The DoLP-weighted mean AoLP is $-14.0^\circ$ to $-14.4^\circ$ at 450~nm (circular spread 3.2--4.0$^\circ$) and $85.0^\circ$ to $85.6^\circ$ at 660~nm (spread 11.6--15.2$^\circ$) for all four devices. The axes at the two wavelengths are nearly orthogonal (Fig.~\ref{fig:disk}), and this wavelength dependence must be recorded as part of the reader configuration. In the polarization maps [Fig.~\ref{fig:setup}(d) and supplementary Fig.~11], the marking appears as structure in $S_0$ and as reduced DoLP, while the AoLP deviation from each device's mean axis, a quantity invisible to $S_0$, carries fine spatial structure that differs between devices.

\subsection{Estimator Fidelity}\label{sec:res-fidelity}
\begin{figure}[!t]
\centering
\includegraphics[width=\columnwidth]{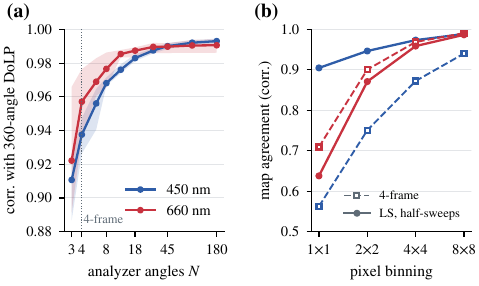}
\caption{DoLP estimator fidelity. (a) Correlation of the $N$-angle DoLP with the 360-angle estimate (device mean as lines, device range as bands). (b) Agreement of the DoLP maps of D1 at both wavelengths vs.\ binning of the full-resolution maps, between four-frame estimates from two shifted angle sets (dashed) and between least-squares (LS) fits on the two half-sweeps (solid).}
\label{fig:fidelity}
\end{figure}

Fig.~\ref{fig:fidelity}(a) compares the DoLP from $N$ evenly spaced analyzer angles with the 360-angle estimate at $134\times256$ resolution, averaged over nine angle offsets. Four angles reproduce the reference DoLP with correlation 0.93--0.95 at 450~nm and 0.94--0.98 at 660~nm, and 12 angles reach at least 0.97. The estimator is, however, strongly resolution dependent [Fig.~\ref{fig:fidelity}(b)]. At the full $538\times1024$ resolution, four-frame DoLP maps computed on $\{0^\circ,45^\circ,90^\circ,135^\circ\}$ and on the same set shifted by $30^\circ$ or $60^\circ$ agree at only 0.56 for D1 at 450~nm, and $2\times2$, $4\times4$ and $8\times8$ binning raises the agreement to 0.75, 0.87 and 0.94, the trend that (\ref{eq:dolpnoise}) predicts for pixel-limited noise. Least-squares fits from the two independent half-sweeps agree at 0.90 at full resolution and 0.99 at $8\times8$. Low agreement between four-frame DoLP maps is therefore an estimator-noise effect, not evidence of weak polarization modulation. At 660~nm the full-resolution half-sweep agreement is lower (0.64, rising to 0.87 with $2\times2$ binning) because the $1\theta$ component differs between the two halves of the sweep. This nuisance enters the verification below as realistic genuine-score degradation.

\subsection{Polarization Separates Devices Better Than Intensity}\label{sec:res-verify}
\begin{figure*}[!t]
\centering
\includegraphics[width=\textwidth]{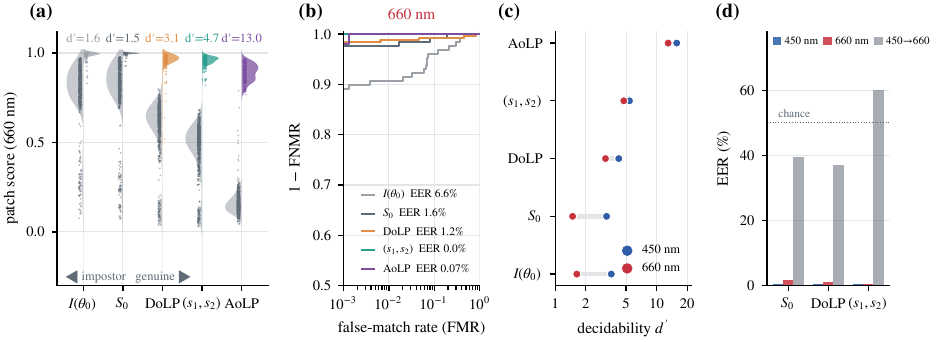}
\caption{Physics-level verification between independent half-sweeps (registered, patch level). (a) Genuine (right) and impostor (left, gray) score distributions at 660~nm for five representations. (b) ROC curves at 660~nm. (c) Decidability $d'$ by representation at both wavelengths (log scale). (d) EER within each wavelength and with enrollment at 450~nm and probe at 660~nm.}
\label{fig:verify}
\end{figure*}

Table~\ref{tab:verify} and Fig.~\ref{fig:verify} report patch-level verification between independent half-sweeps. Genuine scores are high for every representation (mean 0.89--1.00), as expected within one mount. The representations differ in how far impostor scores fall below them. At 660~nm, total intensity separates devices weakly ($d'=1.48$ and EER 1.6\% for $S_0$, $d'=1.63$ and EER 6.6\% for a single frame) because the registered impostor markings correlate at 0.78--0.80. The normalized Stokes vector raises the decidability to $d'=4.74$ with zero EER, and the AoLP field to $d'=13.0$ (impostor correlation 0.16), because the spatial structure of the polarization state differs between nominally identical markings. The same ordering holds at 450~nm, where intensity is already more distinctive ($d'=3.2$) and the polarization representations again exceed it ($d'=5.4$ for $(s_1,s_2)$ and 15.8 for AoLP). At both wavelengths DoLP lies between intensity and the normalized Stokes vector, because as a scalar magnitude it discards the orientation information that carries most of the device-specific structure.

\begin{figure*}[!t]
\centering
\includegraphics[width=\textwidth]{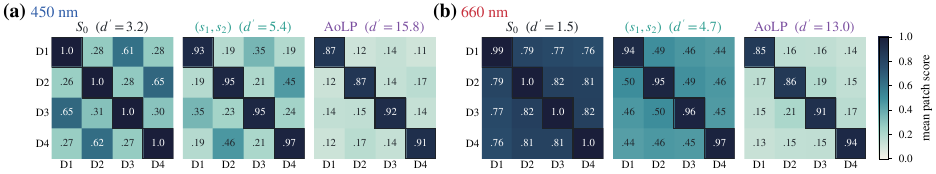}
\caption{Device-by-device verification matrices of the rotation-encoded reader for total intensity $S_0$, the normalized Stokes vector $(s_1,s_2)$ and AoLP at (a) 450~nm and (b) 660~nm. Each entry is the mean registered patch score between the enrollment half-sweep of the row device and the probe half-sweep of the column device (both directions, 64 patch scores per entry). The outlined diagonal holds genuine comparisons.}
\label{fig:vmatrix}
\end{figure*}

Fig.~\ref{fig:vmatrix} resolves these scores by device pair. In intensity, all impostor pairs correlate at 0.76--0.82 at 660~nm, and at 450~nm two pairs (D1/D3 and D2/D4) correlate at 0.61--0.65, so a single intensity threshold must accommodate both similar and dissimilar markings. In AoLP every impostor pair falls to 0.11--0.21 while the genuine diagonal stays at 0.85--0.94.

Enrollment does not transfer across wavelengths [Fig.~\ref{fig:verify}(d)]. Enrolling at 450~nm and probing at 660~nm gives EERs of 39.4\% ($S_0$) and 37.1\% (DoLP), and the normalized Stokes vector falls below chance (AUC 0.38) because the polarization axis rotates by about $80^\circ$ between the wavelengths. A verifier must therefore enroll and verify at the same wavelength, and a two-wavelength fingerprint is best treated as two templates. Fused at score level with (\ref{eq:fusion}), the two templates are complementary. The fused $d'$ rises to 6.73 for $(s_1,s_2)$ (from 5.39 and 4.74), to 4.91 for DoLP (from 4.23 and 3.13) and to 19.5 for AoLP (from 15.8 and 13.0), whereas $S_0$ gains little (3.34 from 3.21 and 1.48) and the fused single-frame intensity has a higher EER (1.2\%) than 450~nm alone (0\%). Both wavelengths come from the same mount, so this gain is a within-session result.

\begin{table}[!t]
\caption{Patch-Level Verification Between Independent Half-Sweeps (256 Genuine / 768 Impostor Scores per Entry)}\label{tab:verify}
\centering\footnotesize
\setlength{\tabcolsep}{3.2pt}
\begin{tabular}{@{}llccccc@{}}
\toprule
$\lambda$ & Representation & $\mu_{\mathrm g}$ & $\mu_{\mathrm i}$ & $d'$ & EER (\%) & AUC\\
\midrule
\multirow{5}{*}{450~nm}
 & $I(\theta_0)$ & 0.991 & 0.373 & 3.57 & 0.0 & 1.000\\
 & $S_0$ & 0.999 & 0.398 & 3.21 & 0.0 & 1.000\\
 & DoLP & 0.971 & 0.334 & 4.23 & 0.0 & 1.000\\
 & $(s_1,s_2)$ & 0.951 & 0.271 & 5.39 & 0.0 & 1.000\\
 & AoLP & 0.890 & 0.139 & 15.78 & 0.0 & 1.000\\
\midrule
\multirow{5}{*}{660~nm}
 & $I(\theta_0)$ & 0.988 & 0.783 & 1.63 & 6.6 & 0.986\\
 & $S_0$ & 0.997 & 0.795 & 1.48 & 1.6 & 0.998\\
 & DoLP & 0.960 & 0.581 & 3.13 & 1.2 & 0.995\\
 & $(s_1,s_2)$ & 0.956 & 0.467 & 4.74 & 0.0 & 1.000\\
 & AoLP & 0.889 & 0.160 & 12.99 & 0.07 & 1.000\\
\midrule
\multirow{3}{*}{450$\to$660}
 & $S_0$ & 0.529 & 0.316 & 0.69 & 39.4 & 0.685\\
 & DoLP & 0.398 & 0.260 & 0.61 & 37.1 & 0.660\\
 & $(s_1,s_2)$ & 0.065 & 0.122 & $-0.44$ & 60.2 & 0.378\\
\bottomrule
\end{tabular}
\end{table}

\subsection{Closed-Set Identification and Leakage}\label{sec:res-ident}
Table~\ref{tab:ident} reports identification under the four protocols (see also supplementary Fig.~12). All protocols reach 0.998--1.000 accuracy at both wavelengths and for both modalities, including the angle-sector protocol in which no test angle or its $180^\circ$ twin is seen in training, so closed-set identification of these four devices is robust to the leakage paths that can be constructed within one session. The corresponding confusion matrices are diagonal except for two misclassified frames, so the device-pair structure is given instead by the verification matrices of Fig.~\ref{fig:vmatrix}.

The controls qualify this accuracy. Random labels give 0.22--0.28 by construction and therefore cannot detect leakage. Labels that mark contiguous angle blocks of a single device are separated with 0.94--0.99 accuracy under a random split, so random frame splits evaluate on near-copies of training frames and any metric computed on them is optimistic. The image mean and standard deviation alone identify the device with 0.51--0.91 accuracy, and an 8-pixel thumbnail reaches 0.79--1.00 on unseen-angle sectors. Coarse layout and illumination differences between mounts thus suffice for closed-set identification of a few devices. An unsupervised embedding of independent 12-angle Stokes estimates (supplementary Fig.~13) likewise separates all devices for both $S_0$ and AoLP (leave-one-out nearest-neighbor accuracy 1.00). Within one session, accuracy and cluster plots therefore do not rank representations, and supervised embeddings separate classes by construction~\cite{mcinnes2018umap}. We consequently base the security claim on the registered impostor separation of Section~\ref{sec:res-verify}.

\begin{table}[!t]
\caption{Closed-Set Identification Accuracy (4 Devices, Chance 0.25)}\label{tab:ident}
\centering\footnotesize
\setlength{\tabcolsep}{3pt}
\begin{tabular}{@{}lcccc@{}}
\toprule
 & \multicolumn{2}{c}{450~nm} & \multicolumn{2}{c}{660~nm}\\
\cmidrule(lr){2-3}\cmidrule(lr){4-5}
Protocol / control & Sweep & DoLP & Sweep & DoLP\\
\midrule
Random frame split & 1.000 & 1.000 & 1.000 & 1.000\\
Contiguous block (prior) & 1.000 & 1.000 & 1.000 & 1.000\\
Block, twins removed & 1.000 & 1.000 & 1.000 & 1.000\\
Unseen-angle sectors & 1.000 & 1.000 & 1.000 & 0.998\\
\midrule
Random labels (null) & 0.278 & 0.226 & 0.222 & 0.242\\
Angle blocks, 1 device (null) & 0.972 & 0.935 & 0.991 & 0.968\\
\midrule
Mean + s.d.\ only & 0.513 & 0.914 & 0.533 & 0.671\\
8-pixel thumbnail & 0.789 & 0.999 & 0.999 & 0.997\\
Brightness-normalized & 1.000 & 1.000 & 1.000 & 1.000\\
\bottomrule
\end{tabular}
\end{table}

\subsection{The Dual-Channel Metalens Reader}\label{sec:res-metalens}
\emph{Library provenance and information content.} Frames sampled across the virtual-angle library (nine per device) match $I_X\cos^2\theta+I_Y\sin^2\theta$ with correlation 1.0000 when built from the contrast-normalized XLP image and the registered YLP image [supplementary Fig.~14(a)]. With the raw YLP capture instead, the agreement decreases toward $90^\circ$ (down to 0.65 for D3) because the raw channel is displaced. As Proposition~\ref{prop:rank} predicts, the first singular component of each mean-subtracted library holds 99.0--99.7\% of its variance, and a nearest-centroid classifier trained on a \emph{single} virtual frame per device ($0.15^\circ$, $45^\circ$ or $89^\circ$) identifies all tested library frames correctly. An unsupervised projection of all 2000 library frames (supplementary Fig.~15) separates the five devices, but each device occupies a one-parameter path rather than a cluster. Identification on the library is thus identification of five XLP/YLP image pairs. We report it as two-channel fingerprinting and treat the library as a visualization of $(S_0,S_1)$, not as additional measurements. With one capture per device, a genuine/impostor analysis of the metalens reader is not yet possible.

\emph{Inter-channel registration.} The two metalenses image the target with an offset that must be corrected in software. Phase correlation between the raw and registered YLP images gives displacements of 0.8--9.5 pixels (supplementary Table~VI), and the raw and registered images agree at only 0.65 for D3. Because $S_1=I_X-I_Y$ is a difference image, a residual misregistration of one pixel produces edge artifacts comparable to the polarization signal itself, which makes registration accuracy a first-order requirement for a dual-channel reader.

\begin{figure*}[!t]
\centering
\includegraphics[width=\textwidth]{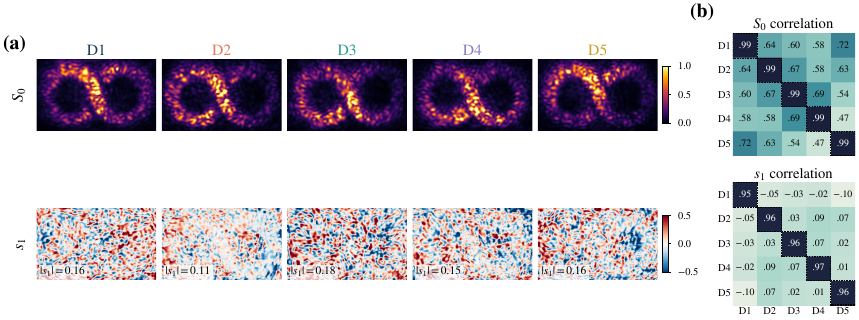}
\caption{Metalens dual-channel measurements at 660~nm. (a) Total intensity $S_0=I_X+I_Y$ (top) and reduced Stokes contrast $s_1=S_1/S_0$ (bottom) for the five devices. Insets give the mean $|s_1|$. (b) Pearson correlation between devices after registration to a common reference for $S_0$ (top) and $s_1$ (bottom). The dashed diagonal correlates interleaved pixel rows of the same image (a split-half reference set by pixel noise, not repeatability).}
\label{fig:mlmaps}
\end{figure*}

\emph{Reduced polarimetric contrast.} The two channels are highly correlated within each device (0.86--0.96), and the reduced contrast $|s_1|$ averages 0.11--0.18 with a near-zero mean of $s_1$ (Fig.~\ref{fig:mlmaps}). By (\ref{eq:reduced}) this would bound the DoLP of the logo from below for gain-matched, noise-free channels, but from single contrast-normalized captures it is not a calibrated DoLP estimate. After registration to a common reference, $S_0$ images of different devices correlate at 0.47--0.72, against 0.99 between interleaved pixel rows of the same image. The $s_1$ field is self-consistent within an image (0.95--0.97 between interleaved rows) and uncorrelated between devices ($|r|\le0.10$), so the polarization field of the metalens images is a spatially structured quantity that differs between captures. With one capture per device, however, it cannot yet be attributed to the device rather than to residual lens and registration effects. Such attribution requires repeated captures and lens swaps (Section~\ref{sec:limits}).

\subsection{How Many Polarization Channels Does a Reader Need?}\label{sec:res-channels}
\begin{figure}[!t]
\centering
\includegraphics[width=\columnwidth]{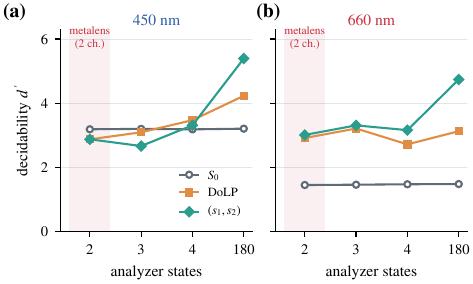}
\caption{Channel-count design study. Decidability of verification between half-sweeps for emulated snapshot polarimeters with 2, 3 and 4 analyzer channels (one frame per channel) and for the 180-angle fit at (a) 450~nm and (b) 660~nm. For the dual-channel design, $(s_1,s_2)$ reduces to $s_1$ and DoLP to the reduced contrast (\ref{eq:reduced}). The shaded column marks the dual-channel configuration of the metalens reader.}
\label{fig:channels}
\end{figure}

From each half-sweep of the reference data we take single frames at the analyzer states of a dual-channel ($0^\circ/90^\circ$, measuring $S_0$ and $S_1$), triple-channel ($0^\circ/60^\circ/120^\circ$) or quad-channel ($0^\circ/45^\circ/90^\circ/135^\circ$) snapshot polarimeter and repeat the verification of Section~\ref{sec:res-verify} (Fig.~\ref{fig:channels}, Table~\ref{tab:readers}). By Lemma~\ref{lem:frame}, the triple- and quad-channel designs are uniform frames with $N=3$ and $N=4$, so their Stokes estimates carry 60 and 45 times the variance of a half-sweep ($N=180$), and the dual design additionally loses $S_2$. The total intensity $S_0=I(0^\circ)+I(90^\circ)$ formed from the same two frames, which is what a conventional camera records, is unaffected by the channel count ($d'\approx3.2$ at 450~nm and $1.5$ at 660~nm). At 660~nm, the design wavelength of the metalens, the dual-channel snapshot reaches $d'=3.0$ for $s_1$, twice the decidability of this total intensity and comparable to the triple- and quad-channel designs (3.2--3.3), and lowers the EER from 3.1\% to 0.5\% without moving parts. At 660~nm the $0^\circ/90^\circ$ axes lie about $5^\circ$ from the mean polarization axis, so $s_1$ captures most of the polarization magnitude but little of its orientation. A $45^\circ/135^\circ$ pair would measure $s_2$ instead. At 450~nm intensity is already error-free on four devices, the snapshot designs (2.7--3.3) do not reach its zero EER, and the polarization advantage appears only in the 180-angle fit ($d'=5.4$ for $(s_1,s_2)$, 15.8 for AoLP). At both wavelengths the gap between the snapshot designs and the 180-angle fit exceeds the gap between two and four channels. Averaging over frames, which by (\ref{eq:var}) reduces per-channel noise, therefore contributes more than measuring $S_2$. For a metasurface reader this favors multi-frame averaging of a dual- or quad-channel snapshot over adding analyzer states.

\begin{table}[!t]
\caption{Verification $d'$ (EER, \%) of Reader Configurations on the Same Devices and Protocol}\label{tab:readers}
\centering\footnotesize
\renewcommand{\arraystretch}{1.1}
\setlength{\tabcolsep}{1.8pt}
\begin{tabular}{@{}lcccc@{}}
\toprule
Reader & States & \shortstack{Moving\\parts} & 450~nm & 660~nm\\
\midrule
Single frame, $I(0^\circ)$ & 1 & no & 3.57 (0.0) & 1.63 (6.6)\\
Total intensity, $S_0$ & 2 & no & 3.19 (0.0) & 1.45 (3.1)\\
Dual-channel snapshot, $s_1$ & 2 & no & 2.88 (7.8) & 3.01 (0.5)\\
Triple-channel snapshot, $(s_1,s_2)$ & 3 & no & 2.67 (13.3) & 3.31 (2.3)\\
Quad-channel snapshot, $(s_1,s_2)$ & 4 & no & 3.31 (6.3) & 3.16 (1.6)\\
Rotating analyzer, $(s_1,s_2)$ & 180 & yes & 5.39 (0.0) & 4.74 (0.0)\\
Rotating analyzer, AoLP & 180 & yes & \textbf{15.78} (0.0) & \textbf{12.99} (0.07)\\
\midrule
Two-wavelength fusion, AoLP & $2\times180$ & yes & \multicolumn{2}{c}{\textbf{19.45} (0.0)}\\
\bottomrule
\end{tabular}
\\[2pt]
\parbox{\columnwidth}{\scriptsize Snapshot readers and the total intensity $S_0=I(0^\circ)+I(90^\circ)$ are emulated with one frame per analyzer state from each half-sweep, with 256 genuine and 768 impostor patch scores per entry.}
\end{table}

\section{Security Analysis and Discussion}\label{sec:discussion}
\subsection{What Carries the Identity}
Three observations locate the device-specific information. First, closed-set identification of a few devices is achievable from coarse acquisition factors alone (Table~\ref{tab:ident}), so accuracy on its own, including the near-perfect accuracy of the preliminary versions~\cite{khalilian2026cleo,khalilian2026escar}, does not demonstrate a surface fingerprint. Second, once markings are registered and compared patch by patch, intensity impostors remain highly correlated at 660~nm, whereas the orientation of the polarization field decorrelates between devices (Table~\ref{tab:verify}, Fig.~\ref{fig:vmatrix}). Third, this orientation structure is invisible to $S_0$ and to DoLP. A fingerprint should therefore be built on the normalized Stokes vector or on AoLP rather than on intensity or the scalar DoLP alone, and evaluated with angle-disjoint splits and registered impostor comparisons.

\subsection{Adversary Considerations}
\emph{Remarking and substitution.} An adversary who reproduces the layout and font of the marking defeats intensity-based identity at the level of the impostor correlations in Table~\ref{tab:verify} (0.78--0.80 at 660~nm), but must also reproduce the local polarization orientation, which correlates at only 0.14--0.16 between nominally identical parts. \emph{Image replay.} Printed or displayed images of an enrolled marking reproduce intensity but generally not the angle-dependent response of the surface. A reader that checks the full Malus response, including its harmonic content and axis, raises the cost of such presentation attacks, which we have not evaluated. \emph{Reader configuration.} Because enrollment does not transfer across wavelengths, the wavelength, source polarization and analyzer orientation are part of the template and must be authenticated with it. \emph{Reader drift.} The $1\theta$ and $4\theta$ components at 660~nm show that imperfections of a rotating-analyzer reader leak into the signature, which snapshot readers avoid at the cost of inter-channel registration.

\section{Limitations and Future Work}\label{sec:limits}
The main limitation is that every device was acquired in one session and one mount, so genuine scores measure repeatability within a mount, and the results establish the \emph{distinctiveness}, not the \emph{reproducibility}, of the polarization signature. The device population is small, with four units for the reference polarimeter, five for the metalens reader and a single marking design in each dataset. Each metalens device was captured once through one lens pair, so device and lens contributions are confounded. The reader also measures only $S_0$ and $S_1$, and its two channels must be gain-matched against an unpolarized reference as a relative gain error biases $s_1$.

Every experiment here converts directly to a cross-session test. For the reference polarimeter, acquisitions after remounting on different days should replace the half-sweep enrollment/probe pair of Section~\ref{sec:protocol}. For the metalens reader, each device should be captured several times with remounting, and at least one device imaged through both lenses or with the array rotated to separate device from lens contributions. Tens of devices from more than one lot would allow EER estimates with meaningful confidence intervals and tests against remarked parts. On the hardware side, the channel-count study motivates a dual- or quad-channel metasurface reader with multi-frame averaging and on-sensor registration, followed by thermal- and illumination-drift tests.

\section{Conclusion}
We characterized the linear polarization response of IC markings as a contactless device signature with a rotation-encoded reference polarimeter at two wavelengths and a dual-channel metalens reader at 660~nm. The marking response is strongly polarized at 450~nm and moderately at 660~nm, and full-sweep least-squares Stokes estimation suppresses the noise that limits four-frame polarimetry. With registered, patch-level comparison between independent half-sweeps, the orientation of the polarization field separates nominally identical devices far more strongly than intensity, enrollment is wavelength specific, and the two wavelengths are complementary. Near-perfect closed-set identification, in contrast, can be reached by coarse acquisition shortcuts, which the angle-disjoint protocols and controls expose. For the metalens reader, the synthesized virtual-angle library is an exact re-parameterization of the two measured channels, and inter-channel registration is a first-order requirement. Emulation on the reference data indicates that at 660~nm a dual-channel snapshot doubles the decidability of the total intensity from the same frames and lowers the EER from 3.1\% to 0.5\%, whereas at 450~nm no snapshot design matches the zero EER of intensity. Cross-session acquisition and a larger device population are the next steps toward a deployable polarimetric authentication primitive.

\section*{Data and Code Availability}
The analysis code, figure scripts, derived results and raw image data are available from the authors on reasonable request.%

\section*{Acknowledgment}
The authors thank the Lurie Nanofabrication Facility at the University of Michigan for support with device fabrication.%

\bibliographystyle{IEEEtran}
\bibliography{refs}

\begin{IEEEbiographynophoto}{Alireza Khalilian}
is with the Department of Electrical and Computer Engineering, University of Michigan--Dearborn, Dearborn, MI, USA, and the Information Systems, Security and Forensics Lab. His research focuses on metasurface and metalens optics, including high-numerical-aperture polarization-selective silicon-rich-nitride metalenses, metalens arrays for compact microscopy, polarization-resolved imaging and nanophotonic sensing, and on their application to contactless hardware authentication.%
\end{IEEEbiographynophoto}
\begin{IEEEbiographynophoto}{Saad Kiyani}
received the M.S. degree in computer and information science from the University of Michigan--Dearborn, Dearborn, MI, USA, in 2026. His thesis, carried out in the Information Systems, Security and Forensics Lab, addressed thermal fingerprinting of embedded devices with machine learning. He is currently a data scientist. His research interests include hardware authentication, device fingerprinting and machine learning.
\end{IEEEbiographynophoto}
\begin{IEEEbiographynophoto}{Saeid Khalilian}
is with the Faculty of Computer Science and Mathematics, University of Passau, Passau, Germany.%
\end{IEEEbiographynophoto}
\begin{IEEEbiographynophoto}{Hasan Ali Arif}
received the M.S.E. degree in electrical and computer engineering from the University of Michigan--Dearborn, Dearborn, MI, USA, in 2026. His thesis, carried out in the Information Systems, Security and Forensics Lab, addressed real-time identification of embedded devices from hardware physical unclonable features in their voltage signals. His research interests include embedded systems, hardware fingerprinting, cybersecurity and machine learning.
\end{IEEEbiographynophoto}
\begin{IEEEbiographynophoto}{Hafiz Malik}
received the B.S. degree from the University of Engineering and Technology and the Ph.D. degree from the University of Illinois at Chicago. He is a Professor of Electrical and Computer Engineering and the Associate Dean for Graduate Education and Research, College of Engineering and Computer Science, University of Michigan--Dearborn, where he leads the Information Systems, Security and Forensics Lab. His research interests include cybersecurity, cyber-physical system and sensor security, multimedia forensics, content authentication and deepfake detection. He received the University of Michigan--Dearborn Distinguished Research Award in 2022 and the College of Engineering and Computer Science Excellence in Research Award in 2020.
\end{IEEEbiographynophoto}
\vfill

\clearpage
\twocolumn[{%
\centering
{\fontsize{22}{26}\selectfont Supplementary Material\par}
\vspace{10pt}
{\large Polarization-Resolved Device Fingerprinting: From Rotation-Encoded Stokes Imaging to a Dual-Channel Metalens Reader\par}
\vspace{8pt}
{\normalsize Alireza~Khalilian, Saad~Kiyani, Saeid~Khalilian, Hasan~Ali~Arif, and~Hafiz~Malik\par}
\vspace{20pt}}]
\markboth{Supplementary Material}{Khalilian \MakeLowercase{\textit{et al.}}: Polarization-Resolved Device Fingerprinting}
\pdfbookmark[0]{Supplementary Material}{supplementary}
\setcounter{section}{0}
\renewcommand{\thesection}{S-\Roman{section}}
\renewcommand{\theHsection}{S\arabic{section}}
\noindent This supplementary material supports the main article. Figures, tables and equations continue the numbering of the main article, so Figs.~\ref{fig:s-maps}--\ref{fig:s-libembed}, Table~\ref{tab:s-metalens} and \eqref{eq:s-var}--\eqref{eq:s-bias} appear here.

\section{Derivation of Proposition 1}
By Lemma~\ref{lem:frame}, the estimates $\hat a,\hat b,\hat c$ are unbiased and uncorrelated with variances $\sigma^2/N$, $2\sigma^2/N$ and $2\sigma^2/N$. Write $r=\sqrt{b^2+c^2}$ and $\phi=\operatorname{atan2}(c,b)=2\psi$, so that $\DoLP=r/a=P$. The gradient of $r$ with respect to $(b,c)$ is the unit vector $(\cos\phi,\sin\phi)$, and that of $\phi$ is $(-\sin\phi,\cos\phi)/r$. Because the covariance of $(\hat b,\hat c)$ is isotropic, first-order propagation gives $\operatorname{var}(\hat r)\approx2\sigma^2/N$ and $\operatorname{var}(\hat\phi)\approx2\sigma^2/(Nr^2)$, with $\hat r$ and $\hat\phi$ uncorrelated. Then
\begin{equation}\label{eq:s-var}
\operatorname{var}(\widehat{\DoLP})\approx\frac{\operatorname{var}(\hat r)}{a^2}+\frac{r^2}{a^4}\operatorname{var}(\hat a)=\frac{\sigma^2}{Na^2}\,(2+P^2),
\end{equation}
and $\operatorname{var}(\hat\psi)=\tfrac14\operatorname{var}(\hat\phi)\approx\sigma^2/(2Na^2P^2)$, which is \eqref{eq:dolpnoise} of the main article. For the bias,
\begin{equation}\label{eq:s-bias}
\mathbb E\big[\hat b^2+\hat c^2\big]=b^2+c^2+\operatorname{var}(\hat b)+\operatorname{var}(\hat c)=r^2+\frac{4\sigma^2}{N}
\end{equation}
holds exactly. For Gaussian noise $\hat a$ is independent of $(\hat b,\hat c)$, and a second-order expansion gives $\mathbb E[\hat a^{-2}]\approx a^{-2}\big(1+3\operatorname{var}(\hat a)/a^2\big)=a^{-2}\big(1+3\sigma^2/(Na^2)\big)$. The product of the two expectations gives \eqref{eq:bias}. The linearization requires $r\gg\sigma\sqrt{2/N}$. Below this level the Rician bias of $\hat r$ dominates.

\section{Per-Device Polarization Maps}
\begin{figure*}[!t]
\centering
\includegraphics[width=\textwidth]{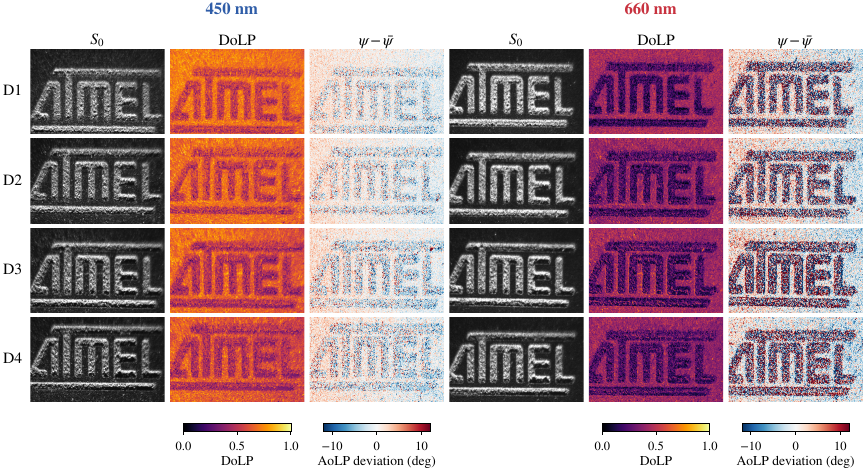}
\caption{Per-device polarization maps from the 360-angle least-squares fit, rotated to landscape. For each wavelength the panels show total intensity $S_0$ (normalized), DoLP and the deviation of AoLP from the device's mean polarization axis.}
\label{fig:s-maps}
\end{figure*}

Fig.~\ref{fig:s-maps} shows the least-squares maps of all four devices at both wavelengths. The marking appears as structure in $S_0$ and as reduced DoLP. The deviation of AoLP from each device's mean axis carries fine spatial structure that differs between devices and is the origin of the high decidability of AoLP in Section~\ref{sec:res-verify} of the main article.

\section{Identification Protocols and Controls}
\begin{figure}[!t]
\centering
\includegraphics[width=\columnwidth]{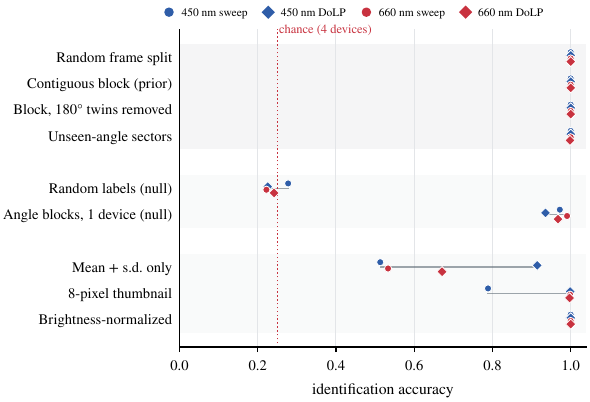}
\caption{Closed-set identification under the four protocols (top band), null controls (middle) and shortcut controls (bottom). Markers encode wavelength and modality.}
\label{fig:s-protocols}
\end{figure}

Fig.~\ref{fig:s-protocols} visualizes the closed-set identification results of Table~\ref{tab:ident} of the main article for the four protocols, the two null controls and the three shortcut controls.

\section{Unsupervised Embedding of the Rotation-Stage Data}
\begin{figure*}[!t]
\centering
\includegraphics[width=\textwidth]{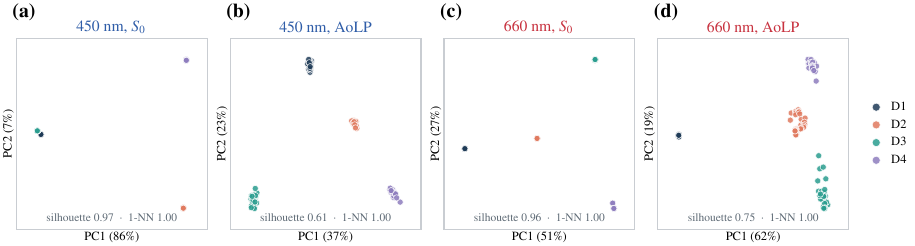}
\caption{Unsupervised embedding of independent 12-angle Stokes estimates (30 per device and wavelength, ECC-registered), projected on the first two principal components of (a), (c) $S_0$ and (b), (d) AoLP encoded as $(\cos2\psi,\sin2\psi)$. Labels are used only for coloring and for silhouette and leave-one-out nearest-neighbor (1-NN) scores, computed on 30 principal components.}
\label{fig:s-embed}
\end{figure*}

Each point in Fig.~\ref{fig:s-embed} is an independent Stokes estimate from 12 analyzer angles spaced $30^\circ$ apart (30 estimates per device and wavelength), registered with the device warps and projected onto the first two principal components without labels. All four devices form separate clusters for both $S_0$ and AoLP (leave-one-out nearest-neighbor accuracy 1.00). The $S_0$ clusters are the tighter ones (silhouette 0.96--0.97 against 0.61--0.75) because a 12-angle AoLP estimate is noisier, as Proposition~\ref{prop:noise} predicts. Within one session, cluster plots therefore do not rank representations. The near coincidence of D1 and D3 in $S_0$ at 450~nm matches their high intensity similarity in Fig.~\ref{fig:vmatrix} of the main article.

\section{Metalens Data Integrity}
\begin{figure*}[!t]
\centering
\includegraphics[width=\textwidth]{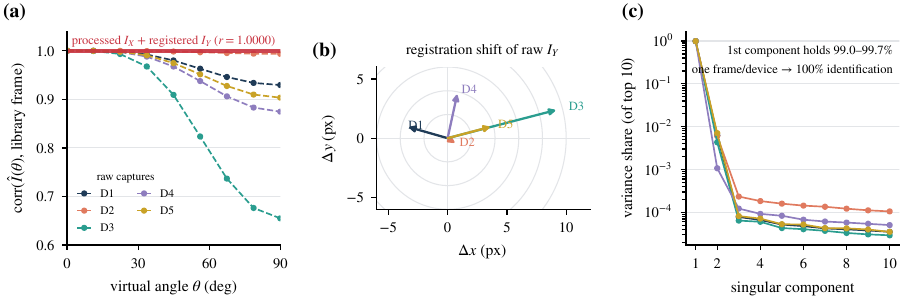}
\caption{Metalens data integrity. (a) Correlation between library frames and the virtual-angle response \eqref{eq:virtual} of the processed $I_X$ and registered $I_Y$ (solid, 1.0000 for all devices) and of the raw captures (dashed). (b) Shift that registration applies to the raw YLP capture. (c) Singular-value spectrum of each mean-subtracted 400-frame library.}
\label{fig:s-mlint}
\end{figure*}

Fig.~\ref{fig:s-mlint} documents the provenance of the virtual-angle library, the displacement between the raw and registered YLP channels and the singular-value spectrum of each library, and Table~\ref{tab:s-metalens} lists the per-device channel statistics.

\begin{table}[!t]
\caption{Per-Device Channel Statistics of the Metalens Reader}\label{tab:s-metalens}
\centering\footnotesize
\setlength{\tabcolsep}{3pt}
\begin{tabular}{@{}lccccc@{}}
\toprule
Device & D1 & D2 & D3 & D4 & D5\\
\midrule
YLP offset (pixels) & 3.5 & 0.8 & 9.5 & 3.9 & 3.8\\
Raw vs.\ registered YLP ($r$) & 0.929 & 0.994 & 0.655 & 0.875 & 0.903\\
XLP vs.\ YLP ($r$) & 0.876 & 0.959 & 0.860 & 0.912 & 0.885\\
Mean $|s_1|$ & 0.163 & 0.110 & 0.182 & 0.150 & 0.160\\
Library 1st component (\%) & 99.2 & 99.0 & 99.4 & 99.7 & 99.1\\
\bottomrule
\end{tabular}
\end{table}

\section{Unsupervised Projection of the Virtual-Angle Library}
\begin{figure}[!t]
\centering
\includegraphics[width=\columnwidth]{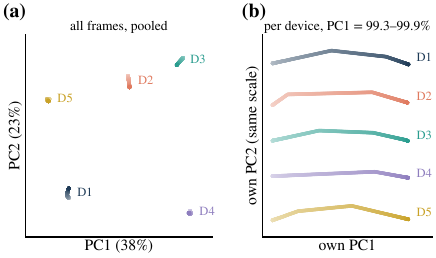}
\caption{Unsupervised projections of the stored virtual-angle libraries (400 frames per device, shaded from light at $\theta=0^\circ$ to dark at $90^\circ$). (a) Principal-component projection of all frames (per-frame standardized). (b) Each device's raw frames in its own first two principal directions, normalized per device with equal axis scales and offset vertically. Each device traces a one-parameter path carrying 99.3--99.9\% of the variance.}
\label{fig:s-libembed}
\end{figure}

Fig.~\ref{fig:s-libembed} projects all 2000 library frames without labels. The five devices separate (leave-one-out 1-NN accuracy 1.00), but each device occupies a short one-parameter path rather than a cluster, and 99.3--99.9\% of each device's library variance lies along a single direction, as stated by Proposition~\ref{prop:rank} of the main article (Table~\ref{tab:s-metalens} gives 99.0--99.7\% because its binned frames were rounded to 8 bits). The residual bending is consistent with the 8-bit quantization and saturation of the stored frames (0.3\% of pixels at 255). An embedding of the library is therefore a picture of five image pairs, not of 2000 observations.

\end{document}